# Electrically tunable VO$_2$-metal metasurface for mid-infrared switching, limiting, and nonlinear isolation

Jonathan King[1*], Chenghao Wan[1,2*], Tae Joon Park[3], Sanket Deshpande[1], Zhen Zhang[3], Shriram Ramanathan[3,4], and Mikhail A. Kats[1,2]

[1]Department of Electrical and Computer Engineering, University of Wisconsin-Madison

[2]Department of Materials Science and Engineering, University of Wisconsin-Madison

[3]School of Materials Engineering, Purdue University, West Lafayette, Indiana

[4]Department of Electrical and Computer Engineering, Rutgers, The State University of New Jersey, New Brunswick, NJ 08854, USA

[*]These authors contributed equally

## Abstract

We demonstrate an electrically controlled metal-VO$_2$ metasurface for the mid-wave infrared that simultaneously functions as a tunable optical switch, an optical limiter with a tunable limiting threshold, and a nonlinear optical isolator with a tunable operating range. The tunability is achieved via Joule heating through the metal comprising the metasurface, resulting in an integrated optoelectronic device. As an optical switch, the device has an experimental transmission ratio of ~100 when varying the bias current. Operating as an optical limiter, we demonstrated tunability of the limiting threshold from 20 mW to 180 mW of incident laser power. Similar degrees of tunability are also achieved for nonlinear optical isolation, which enables asymmetric (nonreciprocal) transmission.

## Introduction

VO$_2$ undergoes a reversible insulator-to-metal phase transition near 68 °C that dramatically changes its optical properties[1]. Because the critical temperature of this phase transition occurs reasonably close to room temperature, there has been considerable recent research in VO$_2$-based optical devices including optical switches[2,3,4], photodetectors[5], smart windows[6], adaptive radiative coolers[7,8,9], thermal camouflage[10,11], and optical memory elements[12,13].

VO$_2$ can be integrated into a variety of structures to enable tunable transmittance and/or reflectance, which can be modulated either by an external control (e.g., via electrical biasing[14] or external heating[15,16]) or by the incident light field itself (like in an optical limiter[17,18,19,20,21,22] or a nonlinear isolator[23,24]). While even a single VO$_2$ film has variable transmittance[25], there is typically a need to improve various device figures of merit, such as the contrast in transmission between VO$_2$ phases, damage threshold, and response time. One approach to improve transmission contrast is to integrate VO$_2$ films or nanostructures into arrays of metal or dielectric optical resonators (often referred to as metasurfaces, though this word has a range of meanings in the literature), such that their resonance frequency and absorption are modified substantially by the phase of the VO$_2$[15,16,18,22,24]. For example, Howes et al. implemented a Mie-type resonator metasurface for optical limiting which enhanced the transmission contrast and power-absorption density within the volume of the VO$_2$, resulting in high speed[18]. Wan et al. demonstrated an optical limiter incorporating a layer of metal aperture antennas on top of a VO$_2$ thin film[22]; this approach also achieved high transmission contrast, but with the metasurface transitioning to a highly reflective, rather than absorptive, final limiting state to enhance the damage threshold of the device. To our knowledge, there has not yet been a demonstration of a single VO$_2$-based integrated device that functions as a limiter or a nonlinear isolator, where the threshold optical power can be dynamically set using an external electrical signal.

Here, we demonstrate an integrated electrically tunable metal-VO$_2$ metasurface that can serve three separate optical functions: switching, limiting, and nonlinear isolation — all of these functions tunable via current applied through the metasurface rather than using an external heater. Our metasurface achieves high optical-transmission contrast by modulating between resonant and non-resonant configurations of aperture antennas, and incorporates Joule heating directly through the metal layer to provide a bias for the limiting and nonlinear isolation functions. For optical switching, current is driven across the metasurface to provide the heat to switch from OPEN (transmitting) to CLOSED (non-transmitting) states. For passive optical limiting, incident light photothermally heats the VO$_2$ layer, driving the device from OPEN to CLOSED. The addition of electric heating enables the tunability of the limiting threshold because it lowers the optical intensity necessary to trigger the transition. Finally, our device also achieves nonlinear optical isolation due to its asymmetric absorptance for forward- and backward-incident light. It thus operates as an optical limiter with two different limiting thresholds depending on which face is illuminated. For incident light of intensity between these two limiting thresholds, nonlinear optical isolation (i.e., asymmetric nonreciprocal transmission of light) is achieved.

By using the metal layer of our metasurface as the current-driven heating element to provide a bias to our optical limiting functionality, we avoid some undesirable consequences that can arise from passing current directly through the VO$_2$. The resistance of VO$_2$ can vary by several orders of magnitude across its transition[26], so Joule heating in VO$_2$ can result in abrupt changes in the current (if driven by a voltage source) while creating localized low-resistance conducting paths (filaments) that generate heat non-uniformly[27,28]. Furthermore, the variation in VO$_2$ resistance across the transition can result in substantial "resistive switching hysteresis" during Joule heating[28,29]. By driving current through the metal layer, our circuit features a comparatively constant resistance, uniform heating, and avoids this resistive switching hysteresis (though the intrinsic hysteresis of the VO$_2$ transition remains). Lastly, by integrating the heater element directly into the metal layer, we add just one small wire-bonding step beyond the general fabrication flow we previously used in VO$_2$-based metasurface optical limiters[22]. Overall, our approach enables high transmission contrast and an elegant low-footprint means of biasing.

**Results**

In designing our multifunctional device, our objectives included the following: 1) Large transmittance contrast between the CLOSED and OPEN states. 2) Large contrast in absorptance of forward-propagating and backward-propagating light in the OPEN state. This absorptance contrast determines the difference in limiting threshold between the two propagation directions and the range of laser power where nonlinear optical isolation can be achieved. 3) An OPEN-state transmittance of ~0.5. Ideally, an OPEN-state transmittance of 1 would be desirable for each function; however, both optical limiting and optical isolation are triggered by absorption in the OPEN state and so some absorption at the expense of transmittance is necessary. 4) A transmittance peak with FWHM of at least 1 μm to accommodate broadband operation in the mid-wave infrared (MWIR). 5) A high-reflectance CLOSED state to enable a high damage threshold for optical limiting and isolation.

We did not rigorously optimize for a single figure of merit, but rather, we sought a design that broadly accommodated these criteria. Our design was inspired by the limiters demonstrated previously[22], comprising resonant aperture antennas in a metal film deposited on an top of a VO$_2$ film. When the VO$_2$ is in the insulating state, the aperture-antenna array features high resonant transmission, which can be reduced and eventually extinguished as the VO$_2$ transitions into its metal state where the array becomes highly reflective. We chose TiO$_2$ as the substrate for VO$_2$ thin-film synthesis, because it is transparent in the MWIR[30] and is lattice-matched with VO$_2$, which enables high-quality VO$_2$ film growth[31]. To roughly optimize our device, we simulated the transmittance and absorbance of this structure in the OPEN (insulating VO$_2$) and CLOSED (metallic VO$_2$) states using the finite-difference time-domain (FDTD) method implemented in Lumerical FDTD, with literature values for refractive indices of gold and TiO$_2$[30,32] and using the refractive index of VO$_2$ that we measured from a test VO$_2$-sapphire sample using mid-infrared variable-angle spectroscopic ellipsometry (see S.I. section 1). In the simulation, the TiO$_2$ substrate was taken to be semi-infinite, and therefore did not account for the backside TiO$_2$-air interface reflectance (~0.15) that is present in the experiment. We tuned our metasurface geometry to achieve a peak CLOSED-state transmittance at 3.9 μm, the operational wavelength of a quantum cascade laser (QCL) in our laboratory.

Our final structure is shown in Fig. 1 and has square coaxial apertures with inner diameter of 450 nm and outer diameter of 650 nm, with unit-cell spacing of 1200 nm. The $VO_2$ thickness is 120 nm. The Au thickness did not have a significant impact on device performance between ~50 to 150 nm; here we present simulations for Au thickness = 80 nm, which corresponds to our fabricated device described below.

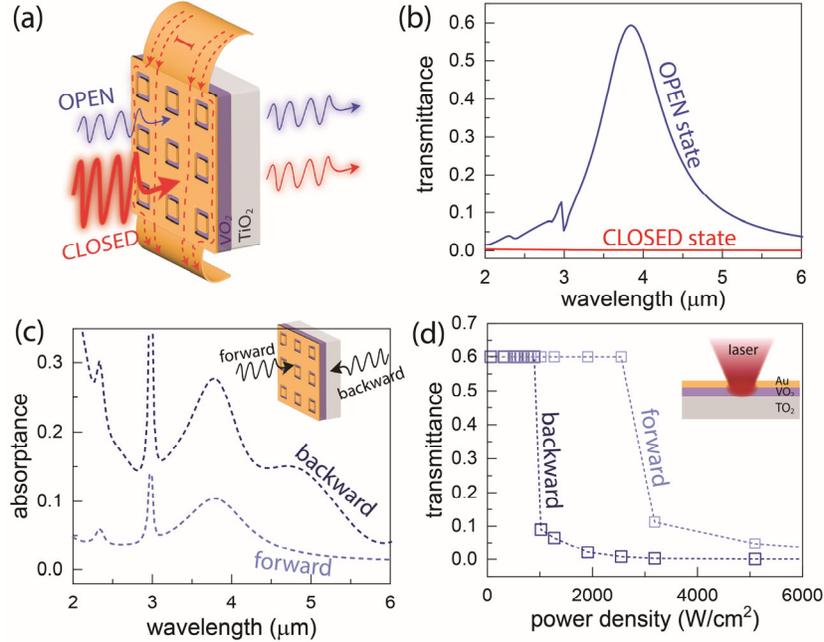

**Figure 1) (a)** Electrically driven device consisting of a gold metasurface, which also serves as our heater, deposited on top of a $VO_2$ thin film on a $TiO_2$ substrate. The CLOSED state can be triggered by electrical Joule heating across the gold surface and/or from photothermal heating from incident light at sufficient intensity. **(b)** Transmittance in OPEN and CLOSED states, simulated using the finite-difference time-domain (FDTD) method. **(c)** Simulated absorptance in the OPEN state for forward- and backward- propagating light. **(d)** Simulated transmittance of our device at $\lambda$ = 3.9 µm as a function of incident intensity for both forward and backward illumination. The coupled thermo-optic behavior was modeled using a combination of FDTD for electromagnetics and FEM for heat transfer. For simplicity, the gold metasurface layer is approximated as a single continuous layer in the FEM heat transfer model since the area density of the apertures is small (see S.I. section 7 for more detail).

To simulate the transmittance for a given incident optical intensity with no electrical biasing [Fig. 1(d)], we used a model that consisted of a transient thermal simulation (implemented in COMSOL Multiphysics) coupled with the FDTD simulation. With no laser irradiation, the device was in thermal equilibrium at room temperature. The transient thermal simulation was initiated by heat flux induced into the device by laser irradiation, based on the FDTD-calculated absorptance at room temperature. The heat flux was Gaussian, to account for the intensity distribution of the laser beam. For each time increment, the thermal simulation returned a transient temperature distribution within the device, which was used to update the absorptance based on the FDTD-simulated temperature-dependent absorption. The temperature-dependent refractive index of $VO_2$ used in our FDTD simulation was generated using the Looyenga mixing rule, an empirical mixing formula that has been previously used for $VO_2$[1]. The resulting absorptance was then fed into the thermal simulation at the next time step. This coupled opto-thermal simulation loop was iterated until the temperature distribution stabilized. Finally, we converted the stabilized temperature distribution to transmittance according to the FDTD transmission results. We used a continuous gold film to represent the gold metasurface in our thermal modeling, which is expected to be a valid assumption because the area density of gold in the metasurface is high (~0.85). The thermal properties of gold, $VO_2$, and $TiO_2$ used in our COMSOL simulation were taken from refs. [33,34,35,36]. This modeling approach is similar to that of our previous works[22,23].

One fabricated metasurface had dimensions of 400 μm × 400 μm. We note that the smallest feature in our metasurface is ~200 nm, and while we fabricated it using electron-beam lithography, it can also be manufactured on a large scale using deep ultraviolent lithography. In addition to the metasurface, gold pads, 200 μm × 400 μm each, were patterned on opposite sides of the metasurface to serve as electrical terminals [Fig. 2(b)]. Our substrate was mounted onto a custom printed circuit board over a 1-cm-diameter through-hole to accommodate transmittance measurements. Each candidate device was Al-wired-bonded to a separate electrical circuit which connected to a Keithley 2612B Source Meter Unit. Additional fabrication details are provided in S.I. section 2.

The transmittance of our wired-up device was measured using an infrared microscope (Hyperion 2000) attached to a Fourier-transform infrared (FTIR) spectrometer (Bruker Vertex 70) [Fig. 2(a)]. Our measurement spot size was smaller than the metasurface patch. Transmittance curves at various current levels (150 mA to 310 mA) are shown in Fig. 2(c), demonstrating an extinction ratio between OPEN and CLOSED states of approximately ~100 for our target wavelength of 3.9 μm. Note that for currents close to the midpoint of the phase transition (180 to 220 mA), the temperature of the $VO_2$ patch and resulting spectrum can take some time to stabilize; see S.I. Section 8.

The OPEN-state transmittance of our device closely resembles that of our simulations [Fig 1(b)]. $VO_2$ exhibits hysteresis across the transition[37], and so we measured transmittance at our target wavelength for both ascending and descending current [Fig. 2(d)]. The hysteresis in current (~15 mA) is much smaller than the current increment needed to switch between states for either individual curve (~100 mA).

We also measured voltage as a function of current in the current-ascending direction through the phase transition [Fig. 2(d)]. Across the transition, $VO_2$ resistance can change by about four orders of magnitude[38,26]. However, the resistance of our biasing circuit stays roughly constant across the transition.

When biased, current preferentially flows through the low-resistance gold layer, regardless of $VO_2$ state, ensuring predictable heating rates. Bias cycling measurements are given in S.I. section 3.

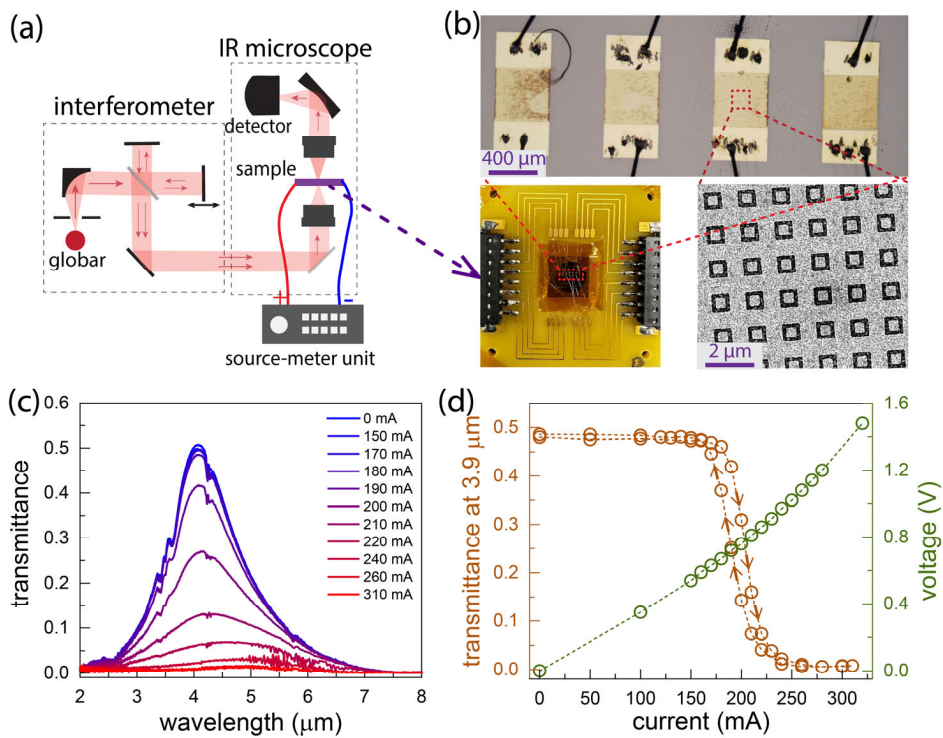

**Figure 2) (a)** Schematic of transmittance measurements. **(b)** Optical-microscope image of the circuit board with four devices, each fabricated with slightly different e-beam dose. The color variations across the metasurfaces are attributed to different local yields of liftoff, as well as some PMMA resist residues. The measurements in the main text all come from the device third from the left. **(c)** Transmittance spectrum of our device for varying current (0 to 310 mA), initially at ambient temperature. **(d)** Transmittance at $\lambda$ = 3.9 μm and voltage, both vs. the current. Transmittance (left axis) is measured for ascending and descending currents.

Next, our device was tested in limiting mode, see Fig. 3(a) for measurement apparatus. Linearly polarized light from a quantum cascade laser (QCL; Alpes Lasers model RT-CW-FP-QCL-HHL) was expanded using a 30 mm concave lens and combined with a 100 mm convex lens before passing through a wire-grid polarizer that served as an attenuator. A second 100 mm convex lens was then used to focus light onto our optical device, achieving a waist radius of ~104 μm (see S.I. section 5). Transmitted light was then collected using a third 100 mm convex lens and measured using a Thorlabs S302C power meter. The temperature of the circuit board was not controlled and was likely close to ambient temperature. For several current levels, we measured the output power vs. input power up to a maximum input power of 200 mW, the limit of our experimental setup [Fig. 3(b)]. For each point on each curve, the current bias was first applied, followed by laser illumination in a power-ascending sequence to ensure that we followed the temperature-ascending side of the hysteresis loop [red points in Fig. 2(d)]. As with the electrical switching measurements, the limiting measurements which resulted in an intermediate $VO_2$ state took some time to stabilize (see S.I. Section 8); note, however, that stabilization time is not to be confused with response time, which we simulate in S.I. section 7.2 to be on the order of 1 to 1000 microseconds, depending on the input power. When varying the bias current, the limiting threshold of our optical limiter is tunable across a range of at least ~200 mW of input laser power.

For a single current value, the limiter experiences three different states: a linear transmitting state, a nonlinear limiting state, and a linear limiting state. As input power increases from 0, the device is initially in its OPEN state and maintains a linear output-vs-input relationship. As the input power increases, the limiter enters a nonlinear limiting state close to the $VO_2$ insulator-to-metal transition. As the input power increases further, the limiter enters an approximately linear output-vs-input relationship where the slope (~.03) approaches the CLOSED state transmittance (~.005). The discrepancy between the slope in the limiting state in Fig. 3(b) and the CLOSED-state transmittance in Fig. 2(c) can be attributed to incomplete transitions near the edge of the beam during laser heating (see S.I. section 7.1).

An "ideal" optical limiter features a high damage threshold (much higher than the limiting threshold), high transmittance for input powers below the limiting threshold (i.e., high "linear transmittance"), and roughly constant output power versus input power for input powers past the limiting threshold[39]. While we did not measure our device to failure, we observe that for a bias of 170 mA, the damage threshold is at minimum larger than 10 times our limiting threshold, based on our measurements. Our limiter, which has low-power linear transmittance of ~0.5, does not achieve constant output power versus input power past the limiting threshold; instead it initially overlimits at the onset of the $VO_2$ phase transition, gradually transitioning to ideal limiting and then underlimiting, ending up in a low-transmittance linear regime at higher powers. We believe that for most sensor-protection applications, such a power out-vs-in profile is suit.

Like other similar $VO_2$-based devices[40], our optical limiter may experience substantial optical bistability, or photothermal switching hysteresis, when heated photothermally (see S.I. section 4). This hysteresis loop is likely caused in part by the dramatic changes in absorptance through the transition, similar to how abrupt changes in electrical resistance can result in pronounced "resistive switching hysteresis" when current is passed through $VO_2$[28,29].

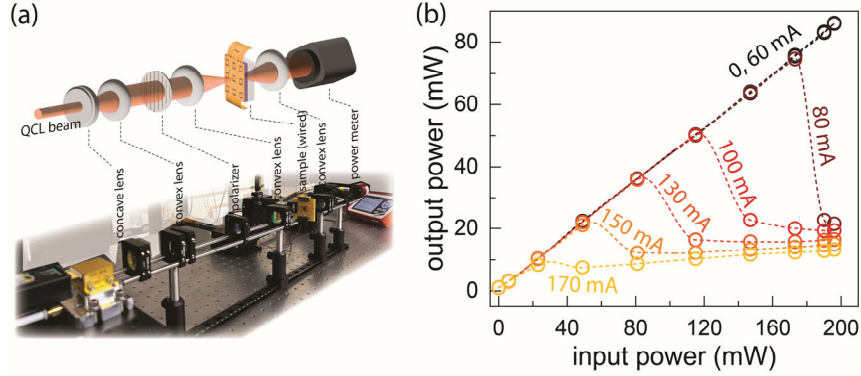

**Figure 3) (a)** Schematic of optical limiting measurement apparatus using a quantum cascade laser (QCL) at 3.9 μm. **(b)** Output-vs-input curves for optical limiting, for different levels of bias current.

Lastly, our device was tested as a nonlinear optical isolator (Fig. 4). Nonlinear isolators achieve asymmetric transmission due to changes in refractive index that depend on incident intensity and direction[41,42]. When our device is in the OPEN state, it is much more absorbing when light is incident from the $VO_2$ side ("backward") vs. the metasurface side ("forward"), with the directions defined in the inset of Fig. 4(a). Therefore, there exists a range of incident intensities for which the device is switched to the CLOSED state for backward incidence but remains in the OPEN state for forward incidence. We used the same setup that was used for the limiting measurements in Fig. 3, and repeated our measurements after reorienting our device such that light illuminated it from the "backward" direction. As the current is increased, the limiting thresholds in both the forward and backward directions decrease, but also the range of intensities where isolation is achieved becomes smaller because electric heating begins to dominate and the optical-absorption asymmetry becomes less significant to the overall heat transfer.

**Discussion**

Our device achieves an isolation ratio of ~5-9 which is a bit larger than existing $VO_2$-based nonlinear isolators operating in the mid-IR[23,24], though this is still much lower than isolation ratios of typical linear isolators based on the Faraday effect (~1000 in the mid-IR). Compared to nonlinear isolators based on conventional nonlinearities, $VO_2$-based devices can operate at much lower optical intensities, especially when thermally biased close to the phase transition. The isolation ratio of our device can be improved by increasing the quality factor of the metasurface at the expense of bandwidth, enhancing the absorption asymmetry (see discussion in Sections 6 and 7 of Supporting Information), and also by sharpening the $VO_2$ phase transition through strain engineering and/or substrate selection[43]. In addition to increasing absorption contrast between backward- and forward-propagating light, we hypothesize that the isolating range could be expanded via a feedback system that alters the electrical bias depending on the incident power.

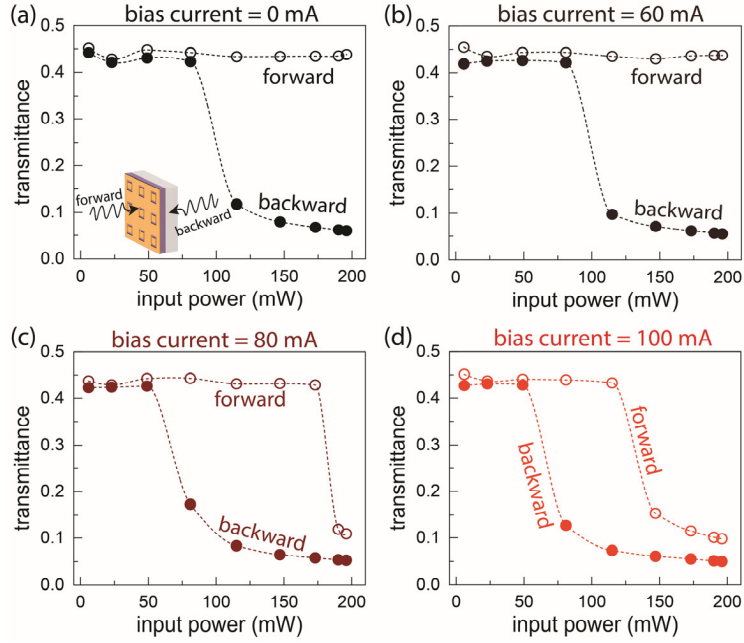

**Figure 4)** Transmittance vs. input power of our device working as a nonlinear isolator (limiting optical diode) for bias currents of **(a)** 0, **(a)** 60 mA, **(a)** 80 mA, and **(d)** 100 mA. As the current is increased, electrical heating begins to dominate over optical heating, reducing the transmittance asymmetry while the limiting threshold in both directions is decreased.

Photothermally driven $VO_2$ switches, limiters, and diodes will generally have a response time which depends on the intensity and pulse shape of the to-be-modulated light[19,20,44]. Intuitively, the response of a can be improved by increasing the volumetric heating rate of the $VO_2$ element per unit input power (e.g., increasing absorption per unit volume), decreasing the heat capacity of the absorbing region, decreasing heat conductance out of the absorbing region by thermally insulating the $VO_2$, and, depending on whether and how the modulator relies on resonance, increasing the quality factor such that resonant transmission is more sensitive to the $VO_2$ transition. Many of these approaches have drawbacks, for example, increasing the quality factor will limit the spectral range of operation, and decreasing heat conductance out of the absorbing element will increase recovery time.

In conclusion, we incorporated Joule heating into a metal-$VO_2$ metasurface to realize a device with three distinct functionalities: optical switching, optical limiting with a tunable threshold, and nonlinear optical isolation. As an optical switch, OPEN and CLOSED states are toggled by varying a DC current from 150 mA to 310 mA. For optical limiting and isolation, the limiting threshold can be varied across tens of milliwatts. Our platform can be further optimized to improve performance in any individual function or for various performance criteria such as transmission contrast between OPEN/CLOSED states or limiter response time or sensitivity. Our approach combining a phase-transition material with an electrically biased metal metasurface can be broadly implemented using related temperature-tunable materials, for example doped[45] or defect-engineered[46] $VO_2$ and rare-earth nickelates[47,48,49].

**Acknowledgements**

MK acknowledges support from ONR (N00014-20-1-2297). SR acknowledges support from AFOSR (FA9550-18-1-0250).

**References**

1. Wan, C. *et al.* On the Optical Properties of Thin-Film Vanadium Dioxide from the Visible to the Far Infrared. *Ann. Phys.* **531**, 1900188 (2019).

2. Guo, Y., Xiong, B., Shuai, Y. & Zhao, J. Thermal driven wavelength-selective optical switch based on magnetic polaritons coupling. *J. Quant. Spectrosc. Radiat. Transf.* **255**, (2020).

3. Qi, J. *et al.* Independent regulation of electrical properties of VO2 for low threshold voltage electro-optic switch applications. *Sens. Actuators Phys.* **335**, 113394 (2022).

4. Guo, Y. *et al.* Dual-band polarized optical switch with opposite thermochromic properties to vanadium dioxide. *Appl. Phys. Lett.* **121**, 201102 (2022).

5. Li, Z. *et al.* Ultrahigh infrared photoresponse from core-shell single-domain-VO2/V2O5 heterostructure in nanobeam. *Adv. Funct. Mater.* **24**, 1821–1830 (2014).

6. Cui, Y. *et al.* Thermochromic VO2 for Energy-Efficient Smart Windows. *Joule* vol. 2 1707–1746 Preprint at https://doi.org/10.1016/j.joule.2018.06.018 (2018).

7. Zhang, W.-W., Qi, H., Sun, A.-T., Ren, Y.-T. & Shi, J.-W. Periodic trapezoidal VO2-Ge multilayer absorber for dynamic radiative cooling. *Opt. Express* **28**, 20609 (2020).

8. Ono, M., Chen, K., Li, W. & Fan, S. Self-adaptive radiative cooling based on phase change materials. *Opt. Express* **26**, A777 (2018).

9. Taylor, S. *et al.* Spectrally-selective vanadium dioxide based tunable metafilm emitter for dynamic radiative cooling. *Sol. Energy Mater. Sol. Cells* **217**, 110739 (2020).

10. Kats, M. A. *et al.* Vanadium dioxide as a natural disordered metamaterial: Perfect thermal emission and large broadband negative differential thermal emittance. *Phys. Rev. X* **3**, 1–7 (2014).

11. Chandra, S., Franklin, D., Cozart, J., Safaei, A. & Chanda, D. Adaptive Multispectral Infrared Camouflage. (2018) doi:10.1021/acsphotonics.8b00972.

12. Jung, Y. *et al.* Integrated Hybrid VO2–Silicon Optical Memory. *ACS Photonics* **9**, 217–223 (2022).

13. Driscoll, T. *et al.* Memory metamaterials. *Science* **325**, 1518–1521 (2009).

14. Ren, Z. *et al.* Active and Smart Terahertz Electro-Optic Modulator Based on VO2 Structure. *ACS Appl. Mater. Interfaces* **14**, 26923–26930 (2022).

15. Tripathi, A. *et al.* Tunable Mie-Resonant Dielectric Metasurfaces Based on VO2 Phase-Transition Materials. *ACS Photonics* **8**, 1206–1213 (2021).

16. Kang, T. *et al.* Mid-infrared active metasurface based on Si/VO2 hybrid meta-atoms. *Photonics Res.* **10**, 373 (2022).

17. Maaza, M. *et al.* Optical limiting in pulsed laser deposited VO2 nanostructures. *Opt. Commun.* **285**, 1190–1193 (2012).

18. Howes, A. *et al.* Optical Limiting Based on Huygens' Metasurfaces. *Nano Lett.* **20**, 4638–4644 (2020).

19. Tognazzi, A. *et al.* Opto-thermal dynamics of thin-film opticallimiters based on the $VO_2$ phase transition. *Opt. Mater. Express* **13**, 41–52 (2022).

20. Guan, H. *et al.* Ultra-High Transmission Broadband Tunable VO2 Optical Limiter. **2200653**, 1–7 (2023).

21. Parra, J. *et al.* Low-threshold power and tunable integrated optical limiter based on an ultracompact VO2/Si waveguide. *APL Photonics* **6**, (2021).

22. Wan, C. *et al.* Ultrathin Broadband Reflective Optical Limiter. *Laser Photonics Rev.* **15**, 1–8 (2021).

23. Wan, C. *et al.* Limiting Optical Diodes Enabled by the Phase Transition of Vanadium Dioxide. *ACS Photonics* **5**, 2688–2692 (2018).

24. Tripathi, A. *et al.* Nonreciprocal optical nonlinear metasurfaces. *arXiv* 3–5 (2022).

25. Béteille, F. & Livage, J. Optical Switching in VO2 Thin Films. *J. Sol-Gel Sci. Technol.* **13**, 915–921 (1998).

26. Kim, H. T. *et al.* Mechanism and observation of Mott transition in VO2-based two- and three-terminal devices. *New J. Phys.* **6**, (2004).

27. Li, D. *et al.* Joule Heating-Induced Metal-Insulator Transition in Epitaxial VO2/TiO2 Devices. *ACS Appl. Mater. Interfaces* **8**, 12908–12914 (2016).

28. Zimmers, A. *et al.* Role of thermal heating on the voltage induced insulator-metal transition in VO2. *Phys. Rev. Lett.* **110**, 1–5 (2013).

29. Murtagh, O., Walls, B. & Shvets, I. V. Controlling the resistive switching hysteresis in VO2 thin films via application of pulsed voltage. *Appl. Phys. Lett.* **117**, (2020).

30. Kischkat, J. *et al.* Mid-infrared optical properties of thin films of aluminum oxide, titanium dioxide, silicon dioxide, aluminum nitride, and silicon nitride. *Appl. Opt.* **51**, 6789–6798 (2012).

31. Kim, H., Charipar, N. A., Figueroa, J., Bingham, N. S. & Piqué, A. Control of metal-insulator transition temperature in VO2 thin films grown on RuO2/TiO2 templates by strain modification Control of metal-insulator transition temperature in VO2 thin films grown on RuO2/TiO2 templates by strain modification. **015302**, (2019).

32. Lide, D. R. *et al. CRC Handbook of Chemistry and Physics*. (2004).

33. Yannopoulos, J. C. *The Extractive Metallurgy of Gold*. (Springer US, 1991). doi:10.1007/978-1-4684-8425-0.

34. Brahlek, M. *et al.* Opportunities in vanadium-based strongly correlated electron systems. *MRS Commun.* **7**, 27–52 (2017).

35. Lee, S., Hippalgaonkar, K., Yang, F. & Hong, J. Anomalously low electronic thermal conductivity in metallic vanadium dioxide. *Science* **355**, 371–374 (2017).

36. Goodfellow. Titanium Dioxide - Titania (TiO2). *AZoM* https://www.azom.com/article.aspx?ArticleID=1179 (2019).

37. Klimov, V. A. *et al.* Hysteresis loop construction for the metal-semiconductor phase transition in vanadium dioxide films. *Tech. Phys.* **47**, 1134–1139 (2002).

38. Hwang, I. H., Park, C. I., Yeo, S., Sun, C. J. & Han, S. W. Decoupling the metal insulator transition and crystal field effects of VO2. *Sci. Rep.* **11**, 1–13 (2021).

39. Van Stryland, E. W., Wu, Y. Y., Hagan, D. J., Soileau, M. J. & Mansour, K. Optical limiting with semiconductors. *J. Opt. Soc. Am. B* **5**, 1980 (1988).

40. Sarangan, A., Ariyawansa, G., Vitebskiy, I. & Anisimov, I. Optical switching performance of thermally oxidized vanadium dioxide with an integrated thin film heater. *Opt. Mater. Express* **11**, 2348 (2021).

41. Kittlaus, E. A., Weigel, P. O. & Jones, W. M. Low-loss nonlinear optical isolators in silicon. *Nat. Photonics* **14**, 338–339 (2020).

42. Shi, Y., Yu, Z. & Fan, S. Limitations of nonlinear optical isolators due to dynamic reciprocity. *Nat. Photonics* **9**, 388–392 (2015).

43. Lee, D. *et al.* Sharpened VO2 Phase Transition via Controlled Release of Epitaxial Strain. *Nano Lett.* **17**, 5614–5619 (2017).

44. Howes, A. *et al.* Optical Limiting Based on Huygens' Metasurfaces. *Nano Lett.* **20**, 4638–4644 (2020).

45. Manning, T. D., Parkin, I. P., Pemble, M. E., Sheel, D. & Vernardou, D. Intelligent Window Coatings: Atmospheric Pressure Chemical Vapor Deposition of Tungsten-Doped Vanadium Dioxide. *Chem. Mater.* **16**, 744–749 (2004).

46. Rensberg, J. *et al.* Active Optical Metasurfaces Based on Defect-Engineered Phase-Transition Materials. *Nano Lett.* **16**, 1050–1055 (2016).

47. Catalano, S. *et al.* Rare-earth nickelates RNiO3: Thin films and heterostructures. *Reports on Progress in Physics* vol. 81 046501 Preprint at https://doi.org/10.1088/1361-6633/aaa37a (2018).

48. Shahsafi, A. *et al.* Temperature-independent thermal radiation. *Proc. Natl. Acad. Sci. U. S. A.* **116**, 26402–26406 (2019).

49. King, J. L. *et al.* Wavelength-by-Wavelength Temperature-Independent Thermal Radiation Utilizing an Insulator-Metal Transition. *ACS Photonics* **9**, 2742–2747 (2022).

Supporting information

# Electrically tunable VO₂-metal metasurface for mid-infrared switching, limiting, and nonlinear isolation

Jonathan King[1*], Chenghao Wan[1,2*], Tae Joon Park[3], Sanket Despande[1], Zhen Zhang[3], Shriram Ramanathan[3,4], and Mikhail A. Kats[1,2]

[1]Department of Electrical and Computer Engineering, University of Wisconsin-Madison

[2]Department of Materials Science and Engineering, University of Wisconsin-Madison

[3]School of Materials Engineering, Purdue University, West Lafayette, Indiana

[4]Department of Electrical and Computer Engineering, Rutgers, The State University of New Jersey, New Brunswick, NJ 08854, USA

[*]These authors contributed equally

1. **Refractive indices of thin film VO₂ on TiO₂ substrate**

We determined the refractive index of VO₂ by performing variable-angle spectroscopic ellipsometry on ~100-nm VO₂ grown on a sapphire substrate, using identical synthesis conditions as our samples on which we fabricated devices, which are on TiO₂ substrates. We used a J. A. Woollam IR-VASE Mark II ellipsometer. In our ellipsometry modeling, we used a series of Lorentz oscillators to describe the insulating-phase dielectric function of VO₂ and an additional Drude term to capture the contribution of the free carriers for the metallic phase. More details can be found in Ref. [S1]. The real and imaginary parts of the refractive index are plotted in Fig. S1.

The reason we used VO₂ films on sapphire rather than on TiO₂ for the ellipsometry measurements is that our substrates were double-side polished (to enable the transmission measurements in the main text), and had to be backside-roughened to enable infrared ellipsometry measurements. However, TiO₂ is soft and brittle, and we had a hard time roughening the backsides without breaking the samples. As a result, we measured the refractive index by analyzing comparable films grown on a sapphire substrate. While we can expect some variations in optical properties of VO₂ films deposited on different substrates, the differences should be very minor in the 2-11 μm range[S1], which covers the measurements in this paper.

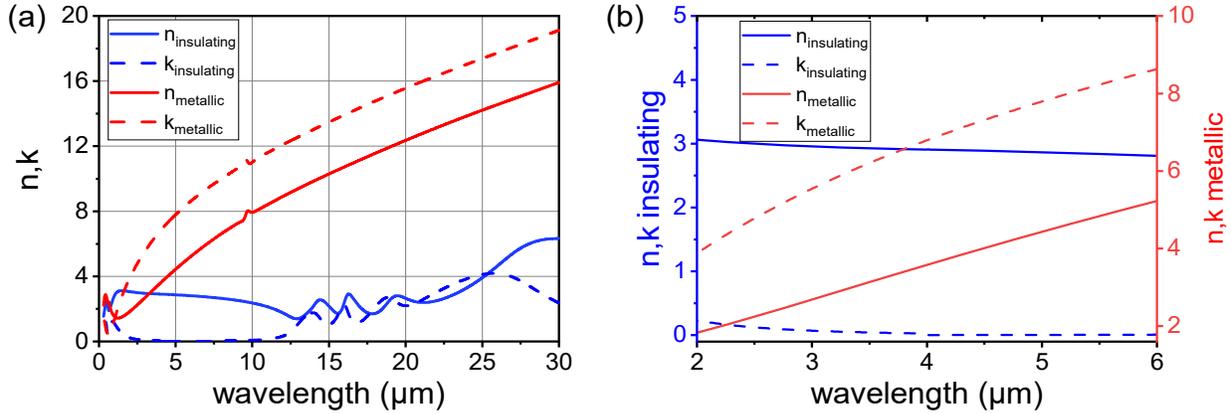

**Figure S1)** Complex refractive index of $VO_2$ on sapphire in insulating and metallic states obtained by spectroscopic ellipsometry. (a) Fit over entire fitting range (b) Zoom-in into the 2-6 micron wavelength range.

## 2. Device Fabrication

We purchased $TiO_2$ substrates from MTI Corp. $VO_2$ thin films with a thickness of 100 nm were deposited on $TiO_2$ substrates via AJA UHV magnetron sputtering. The background pressure of the deposition chamber was pumped down to $9 \times 10^{-7}$ Torr. The working pressure of the deposition chamber was kept to $5 \times 10^{-3}$ Torr with flowing the mixture of Ar gas (49.5 sccm) and $O_2$ gas (0.5 sccm). With RF sputtering of $V_2O_5$ ceramic targets, the deposition of $VO_2$ thin films was performed at 750 °C. PMMA (495 PMMA A4) was spin coated onto our $TiO_2$-$VO_2$ thin film square and our metasurface pattern was written using an Elionix GS-100 e-beam lithography system. After exposure, we developed our square in a MIBK/IPA solution (volume ratio of 1/3), and then we deposited a 5 nm layer of chromium to assist with adhesion followed by 80 nm of gold using metal evaporation. For our liftoff procedure, we first soaked the thin film square for 25 minutes in a large, loosely-covered, beaker of acetone heated to 65 °C. Then, we sonicated the acetone beaker (unheated) for 75 seconds. We transferred the square to a "clean" acetone bath in a separate beaker, making sure to transfer quickly enough to prevent the acetone film on the square surface from evaporating during transfer. We waited for 5 minutes before sonicating the second beaker for another 75 seconds. We finished with a quick 1-minute IPA bath and rinse before blow-drying the square with an air gun.

We imaged our device using a Nikon Eclipse LV150/LV150A microscope at 20× and 100× magnifications (Fig. S2). Based on these images, we estimate a fabrication yield of 85-95% of the apertures, with defects primarily being failure to lift off. We expect yield can be significantly improved following further refinement of the fabrication process.

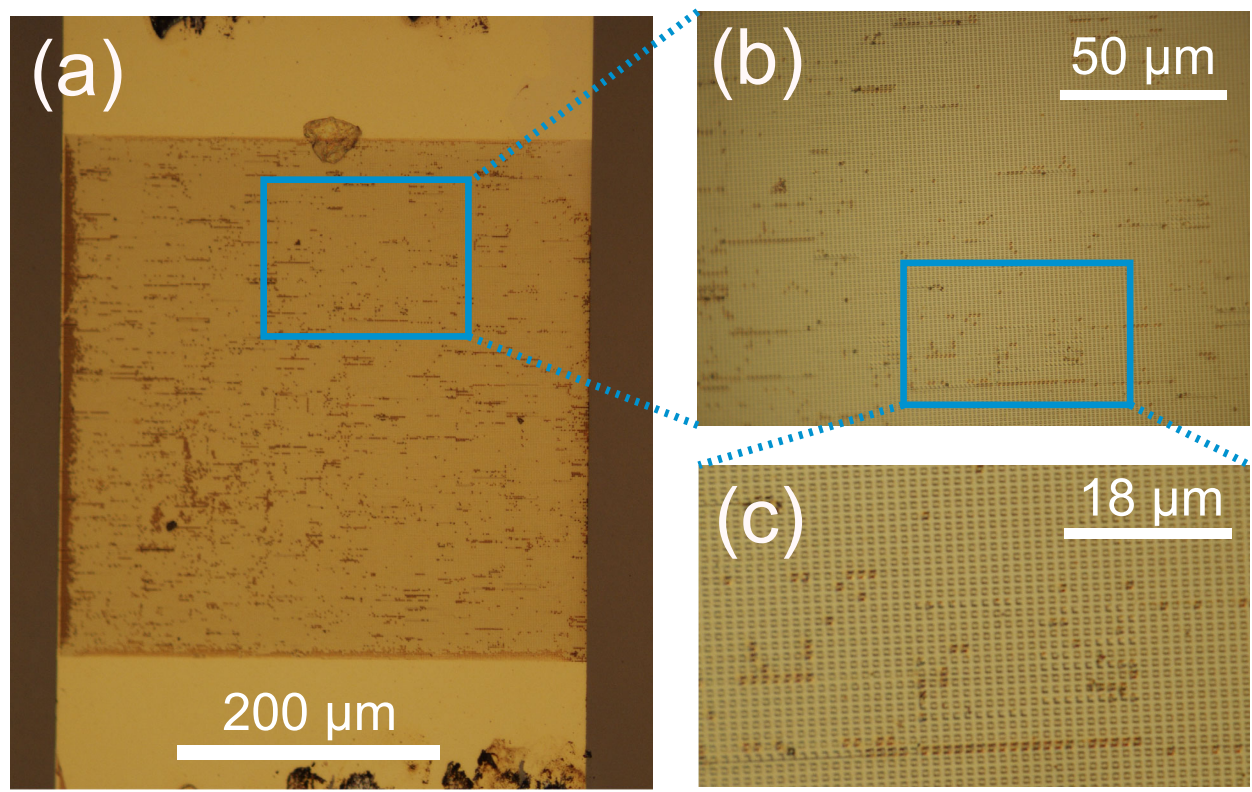

**Figure S2)** Microscope images of device at magnifications of (a) 20×, (b) 100×, (c) 100× (zoomed-in)

## 3. Electric bias cycling

After switching, limiting, and isolating experiments were completed, we attempted to conduct electric cycling experiments on our device, however, while handling our PCB prior to cycling measurements, one of the wire bonds detached from the gold contact pad of our device. Instead of using our original device, we used the leftmost device pictured in Fig. 2(b) in the main text, as shown in Fig. S3 below.

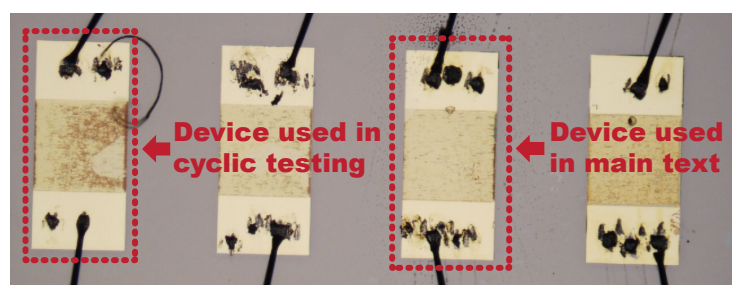

**Figure S3)** Optical-microscope image of the circuit board with four devices, each fabricated with slightly different e-beam dose, also shown in Fig. 2(b) of the main text. The color variations across the metasurfaces are attributed to different local yields of liftoff, as well as some PMMA resist residues. The device used in cyclic testing is the leftmost device, the device used for all other measurements in both the main text and the supplementary is the third from the left.

Our cycling experiments consisted of cycling from 0 mA to 310 mA and back to 0 mA for a total of 23 cycles, pausing for about a minute for each current. We tracked transmittance at our target $\lambda = 3.9$ μm for the first 12 cycles, results shown in Fig. S4. Transmittances appear to be stable through our cycling.

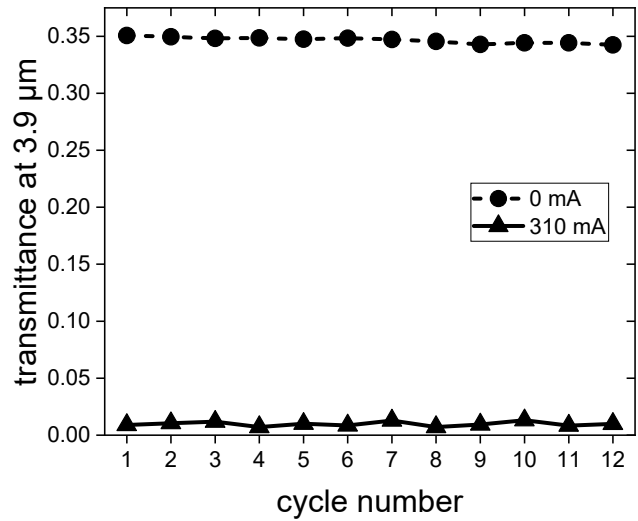

**Figure S4)** Transmittance at $\lambda = 3.9$ μm for 12 cycles from 0 mA to 310 mA and back.

For cycles 1, 12, and 23, we stopped to measure the transmittance for various current values in both the current-ascending and current-descending directions (Fig. S5). For intermediate currents between 0 mA and 310 mA, we measured multiple times until consecutive measurements, taken approximately one minute apart, yielded values that were within the measurement errors of our setup. After the first cycle or so, the hysteresis curves appear consistent in both the current-ascending and current-descending directions. We do not know for sure why the first cycle has a transition at apparently higher current, but we believe it is due to the electrical contact being conditioned.

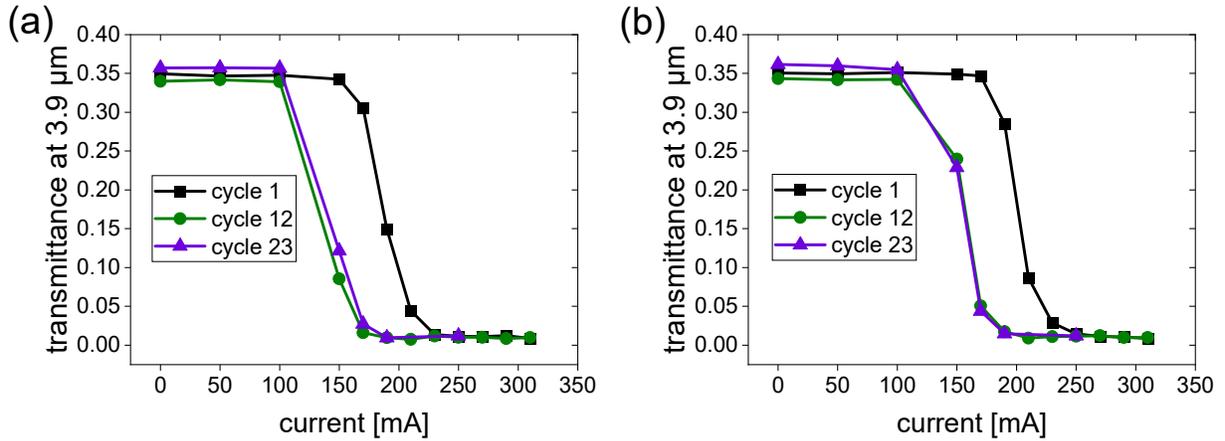

**Figure S5)** Transmittance at λ = 3.9 µm for cycles 1, 12, and 23 in (a) current-ascending and (b) current-descending directions.

## 4. Photothermal switching hysteresis

In the main text in Fig. 3(b) and Fig. 4, we only plotted data for increasing input optical power. In Fig. S6, we include the curve for decreasing input optical power. Specifically, we are looking at the 80 mA electrical bias in both the forward and backward incidence cases.

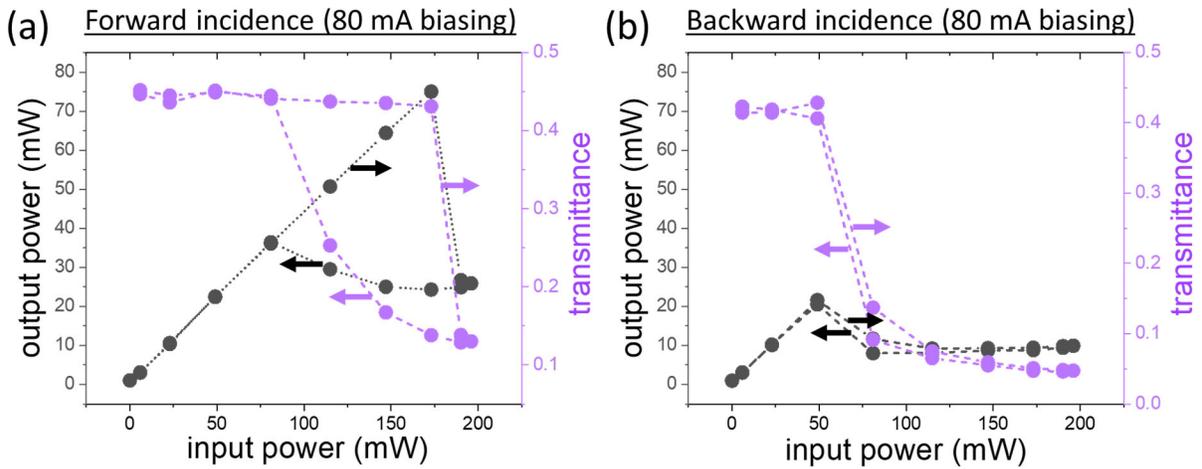

**Figure S6) (a)** Output power and transmittance vs. input power in the ascending-power and descending-power directions for the forward-incident configuration of our modulator (λ ~ 3.9 µm). Pronounced hysteresis occurs between the ascending-power and descending-power directions presumably due to macroscopic photothermal hysteresis effects in addition to the intrinsic hysteresis of $VO_2$. **(b)** Output power and transmittance vs. input power in the ascending-power and descending-power directions for the backward-incident configuration of our modulator. Hysteresis is much more subdued and is presumably limited to the intrinsic hysteresis in $VO_2$.
55

In Fig. S2(a), we observe what appears to be optical bistability—a large hysteresis vs. input optical power that cannot be simply attributed to the intrinsic temperature hysteresis in $VO_2$[S2]. We explain this effect with the following toy model.

*Toy model*

A photothermally-heated object that experiences an abrupt increase in optical absorptance vs. temperature may experience excessive hysteresis with respect to incident power, even if the constituent materials of the object experience no intrinsic temperature hysteresis. In Fig. S7, we present a hypothetical object that has optical absorptance that abruptly changes with temperature and has no intrinsic temperature hysteresis. However, there can exist different steady-state absorptances and temperatures for the same input power [e.g., power level 3 in Fig. S7(a) and Fig. S7(b)]. As the optical power increases to level 3, the object still has a low absorptance and so it takes additional input power to generate the heat necessary to reach the transition temperature. Meanwhile, if the object starts out at high temperature due to high incident power, and then the incident power is reduced to level 3, the absorptance remains high, so it takes an additional reduction in input power for the temperature to drop below the transition point. As a result, this object features significant hysteresis when photothermally heated even though it has no temperature hysteresis.

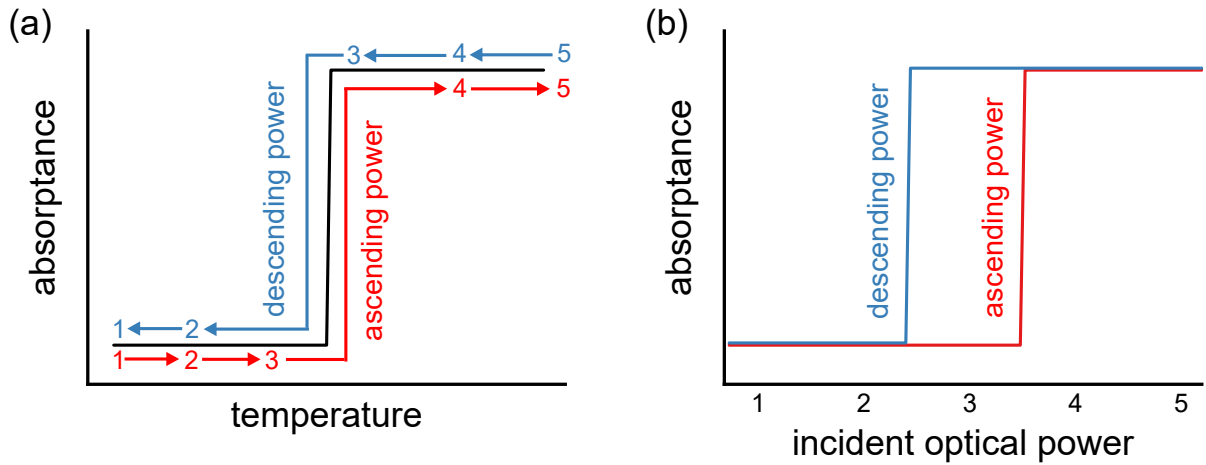

**Figure S7) (a)** Absorptance vs. temperature for a hypothetical photothermally heated object with an abrupt change in absorption at some transition temperature. Each number (i.e. 1, 2,…, 5) corresponds to a different incident optical power level. In the ascending power case, more incident power is necessary to reach the transition temperature because absorptance starts lower. **(b)** Absorptance vs. incident optical power for the same photothermally heated object. There exists pronounced hysteresis even though there is no intrinsic hysteresis with respect to temperature.

Fig. S8 provides the absorptance and transmittance of our device across $VO_2$ phase volume fraction, calculated using FDTD with the same refractive indices that we used in the FDTD calculations in the main text. In addition to exhibiting temperature hysteresis, we note that our device features abrupt increases and non-monotonic changes in absorptance with respect to phase volume fraction. It is therefore potentially susceptible to photothermal switching hysteresis.

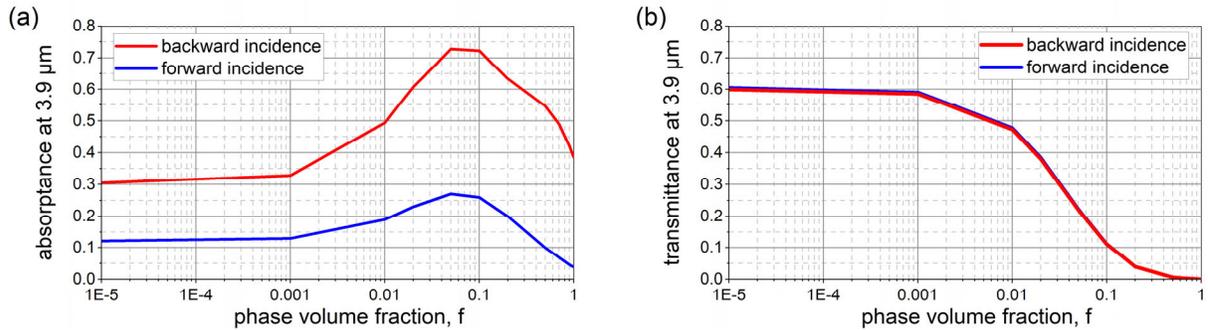

**Figure S8) (a)** Absorptance of our device at $\lambda$ = 3.9 µm vs. the $VO_2$ phase volume fraction, calculated using Lumerical FDTD and the Looyenga mixing rule. **(b)** Optical transmittance vs. phase volume fraction of our modulator at $\lambda$ = 3.9 µm.

This effect is complicated, and full understanding requires accounting for several factors beyond our toy model which may include temperature-dependent Joule heating, spatially variant photothermal heating, latent heat, and the intrinsic temperature hysteresis of $VO_2$. The toy model does not capture why we observe photothermal switching hysteresis in one direction but not the other in Fig. S6. We note that a potentially analogous hysteretic effect, "resistive switching hysteresis", is present for $VO_2$ devices that experience abrupt changes in resistance vs. temperature where substantial hysteresis can occur between current-ascending and current-descending paths[S3,S4].

## 5. Characterization of focused beam size and optical alignment

We characterized the focused beam by placing a circular aperture of radius 125 $\mu m$ at the focal point of the 2f telescope system constructed using the convex lenses [Fig. 3(a) in the main text]. Using the measurements from the power meter, we observed that this aperture clips ~5.5% of the incident power. Assuming a Gaussian beam profile, we calculated the $1/e^2$ waist radius of the focused beam to be ~104 $\mu m$ and the corresponding full width half maximum (FWHM) to be ~122 $\mu m$.

After characterizing the beam at the focal point, we carefully aligned the device, such that our metasurface was positioned at the focal spot of the beam in the 2f telescope system. The alignment process was broken down into two steps: longitudinal alignment and transverse alignment. We mounted the device containing the metasurfaces on a 3-axis, motorized translation stage. The longitudinal direction is labeled as the z-axis of the stage and the two, orthogonal transverse directions (parallel and perpendicular to the optical table) are labeled as x-axis and y-axis. Note that a similar process was followed to align the aperture for characterizing the beam profile (described above).

For the first alignment step, we mount the device such that the beam is approximately incident on one of metasurfaces (without much transverse alignment) and place it between the two convex lenses. We conducted the alignment by moving the device along the longitudinal direction of the beam while observing the readout of the power meter. As the device was moved closer to the focal spot, the power meter reading would reduce as the $400 \times 400\ \mu m^2$ metasurface captured a greater portion of the beam (FWHM ~ ~122 $\mu m$ at focus) before eventually plateauing. We used the central point of the plateau region as an approximate location of the focal spot.

For the second alignment step, we aligned the metasurface in the transverse directions (x and y axes). This was done by moving the translational stage with small increments in each axis and locating the point of minimum transmission (lowest power meter reading) along one direction, and maximum along the other. The asymmetry is because to the left and right of the metasurface there is $VO_2/TiO_2$ with higher transmission, and to the top and bottom there are the opaque gold pads. This process is repeated consecutively several times for each axis to ensure that the beam center overlaps with the metasurface center.

In Fig. S9, we plot the transmittance vs. current (where current is ascending) for our laser setup at low laser power (5.9 mW), and note that even at the highest current (350 mA), the measured transmittance (0.03) does not reach the expected transmittance in the metal state based on the microscope-based FTIR measurement in Fig. 2(c) in the main text. If we had perfect alignment and a perfect gaussian beam shape in the measurement for Fig. S5, more than 99.94% of the incident power should be incident within the bounds of our 400 $\mu m \times$ 400 $\mu m$ metasurface patch. However, we do not expect a perfect gaussian profile nor perfect alignment and we conjecture that a small but non-trivial portion of the incident light avoids the metasurface and instead strikes the unpatterned, Au-free, $VO_2$-$TiO_2$ just outside the metasurface.

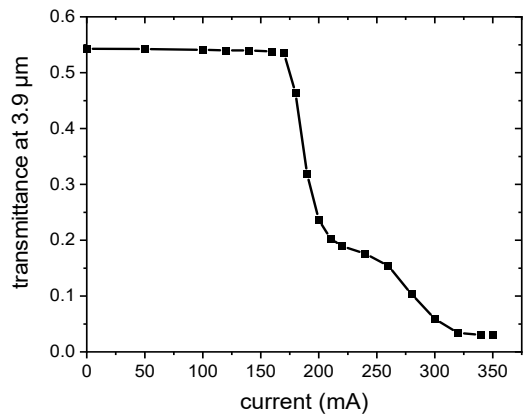

**Figure S9)** Transmittance at $\lambda$ = 3.9 μm vs. current of our device measured using our laser setup, current is ramped in the ascending direction.

## 6. Tradeoff between absorption asymmetry and transmittance

One can optimize the absorption asymmetry via tuning the geometries of the Au and $VO_2$ films, thus enhancing the working power range for the optical isolation function. However, while optimizing absorption asymmetry, it's important to consider that an increase in absorption often results in a trade-off with transmission. The challenge lies in finding a balance that suits specific applications.

Here, we altered the thickness of the Au film compared to the design in the main text as a simple example of this tradeoff. Figure S10(a) shows that when the Au thickness is decreased to half of that of the original design, the transmittance can be increased to 0.7, but at the cost of a reduced absorption asymmetry to ~2. Conversely, if we increase the Au thickness to 250 nm, we can achieve an absorption asymmetry of >5, but this would result in a decrease in transmittance to 0.3, as shown in Figure S10(c).

In this paper, we selected a 70 nm thickness for the Au film, offering a good balance between transmittance and absorption. Note that due to the limited laser power used in our experiments, we required the absorption not to be negligible for photothermal triggering. For applications interacting with much higher laser powers, alternatives such as dielectric metasurfaces or dielectric multilayered structures can be considered, which could exhibit lower absorption, leading to higher transmittance.

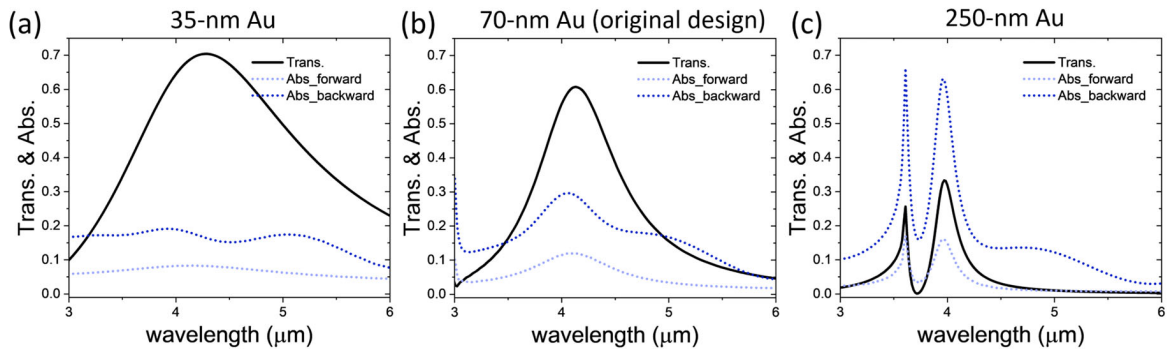

**Figure S10. (a)** Thinner Au film will result in higher transmittance at the cost of a reduced absorption asymmetry. **(b)** The design that is used in the main text. **(c)** Thicker Aul film will result in lower transmission but a higher absorption asymmetry. Note that when altering the Au thickness, we needed to slightly modify the width of the Au apertures and the periodicity to keep the transmission peak around 4 $\mu$m.

## 7. Optothermal simulation

### 7.1 Steady-state simulation

We built a photothermal simulation that couples the optical simulation (FDTD) with a heat-transfer simulation (Heat Transfer Module in COMSOL Multiphysics). Our axisymmetric model consists of five domains centered on the center axis of the laser beam [Figure S11(a)]. The heat generation within the device is due to laser light absorbed in the gold and VO$_2$ layers [domains 1 and 3 in Figure S11(a)]. The laser beam used in our model had a Gaussian-like power distribution with a beam waist radius of 105 μm, making it comparable with our experimental conditions, as discussed in Section 5. We used a continuous gold film to represent the gold FSS in our thermal modeling, which is expected to be a valid assumption because the area density of gold in the FSS is high (~0.85) and the apertures are small. The thermal constants of gold, VO$_2$, and TiO$_2$ were taken from Refs. [S5,S6,S7,S8]. Note that the thermal properties of VO$_2$ were taken from its insulating phase, but their changes due to the IMT are not dramatic (less than 20% across the IMT[S7]), so for simplicity we set them to be constant. This optothermal simulation setup is similar to what we used in Refs. [S9,S10], though in those references a flat-top beam was assumed whereas here we assume a Gaussian intensity distribution to better match our experiment.

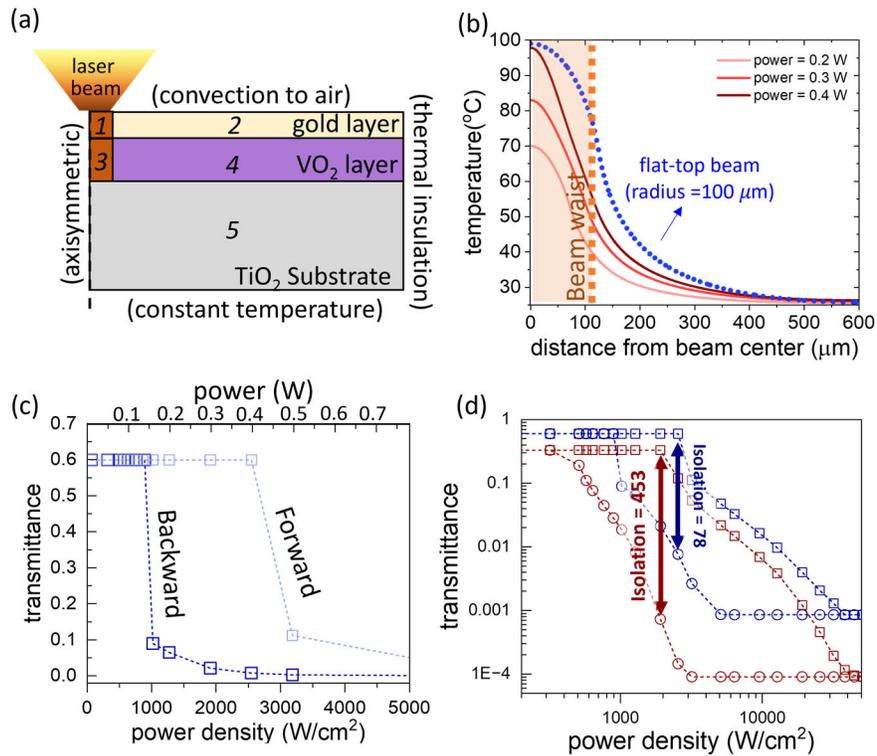

**Figure S11.** **(a)** Schematic of our heat-transfer model implemented in COMSOL Multiphysics. The simulation has axial symmetry (i.e., regions 1 and 3 are at the center of the device). **(b)** Simulated radial temperature distribution upon illumination by a Gaussian beam at backwards incidence, with incident powers of 0.2, 0.3, and 0.4 W. As a comparison, we did a simulation using flat-top beam with a radius of 100 μm, as shown by the blue curve. **(c)** Simulated forward and backward transmittance for different incident powers. **(d)** Comparison of isolation bandwidth and ratio between two designs shown in Figure S6(b and c), which are our original design and a design featuring higher absorption asymmetry, respectively.

Our model works as follows to calculate the transmittance given a particular incident power for a given bias temperature (i.e., the temperature of the bottom of the substrate, which is 25 °C in the simulations in Fig.

S11). The transient thermal simulation is initiated by heat flux into the device by laser irradiation. For each increment in time, the simulation returns a transient temperature distribution within the device, which is used to update the optical absorptance based on the measured temperature-dependent absorption in Figure S4(a). The resulting absorptance is then fed into the simulation at the next time step. This coupled opto-thermal simulation loop iterates until the temperature distribution stabilizes. Finally, we can convert the stabilized temperature distribution to transmittance according to Figure S8(b).

The simulated triggering power and transmission asymmetry (i.e., isolation ratio) [Figure S11(c)] show reasonable agreement with experimental data [Figure 4(a) of the main text]. For example, our simulation predicts that with backward incidence and zero bias current, the device will be triggered at ~0.16 W, compared to ~0.11 W in the experiment. This slight difference is likely due to (a) Fabrication errors in the thicknesses of gold and $VO_2$, which can significantly impact the laser power density within these films, and (b) A subtle mismatch between assumed and actual values of the thermal and optical constants.

To test the influence that small changes in optical and thermal properties and experimental parameters have on the simulated outputs, we performed a test where we decreased the thermal conductivity of $TiO_2$ by 10% from the literature value, and reduced the laser waist by 10% from our best estimate in our experiment. Following these adjustments, our simulation predicted that the triggering power for backward incidence would decrease to ~0.11 W which matches our experimental results [as shown by the red curve in Figure S12(a)].

Our simulation [Figure S11(b)] predicts a difference of tens of degrees in temperature within the beam waist. This implies that for laser powers such that the isolator is just triggered (i.e., from 0.2 to 0.4 W for backward incidence in Fig. S11), the $VO_2$ in the central region of the beam is in the metallic phase, while the region close to the edge of beam waist remains in the insulating phase. This leads to a modest isolation ratio in this low-power regime; specifically, a simulated isolation ratio of 5 – 15 when the incident power increases from 0.2 to 0.4 W. These simulations align well with our experimental observation where the isolation ratio was only 5 – 9 when triggered by backward incidence, as shown in Figure 4(a) of the main text.

For comparison, we simulated a flat-top beam with a 100-μm radius and an incident power of 0.47 W, resulting in a temperature of ~100 °C at the beam center. The resulting temperature profile [blue curve in Figure S11(b)] has a significantly reduced temperature gradient within the beam radius. This simulation helps explain the higher transmission contrast (~100) observed in the optical switching function, characterized by our FTIR microscope [i.e., results in Figure 2(c, d) in the main text]. In our FTIR measurements, the entire 400-by-400-μm Au metasurface area was heated uniformly due to the applied current, which should result in a similar temperature distribution as if the sample were illuminated via a flat-top beam with a 200 μm radius. Moreover, the spot size in our FTIR measurements was ~200 μm on a side, and the measurement was in the center of the device. Within this detection area, our simulations suggest a temperature difference of < 5 °C. Therefore, under the highest bias currents in our measurements in Fig. 2(c), the $VO_2$ should be fully transitioned to the metallic phase, thereby resulting in the high transmission contrast between the OPEN and CLOSED states.

Increasing the absorption asymmetry can serve as one method to enhance both the isolation power range and ratio, though as discussed in the previous section, this could come at the cost of a lower OPEN-state transmission. As a demonstration, we performed an intensity-dependent simulation similar to the one that generated Figure S11(c), but for the design shown in Figure 10(c) which exhibits a higher (i.e., 4-fold) absorption asymmetry. We then compared these results with those from our original design. As shown in Figure S11(d), the higher absorption asymmetry did contribute to an extension of the isolation power range, accompanied by a significantly enhanced isolation ratio of ~450.

7.2 Transient simulation

Our optothermal modeling can be used to calculate the response time of the device. The transient transmittance can be obtained by converting the temperature distribution at each time step to the transmittance, and the response time can then be estimated by finding the time needed to reduce transmittance by $e^{-1}$.

Taking the backward incidence in our modified simulation [i.e., the red curve in Figure S12(a), previously discussed in Section 7.1] as an example, our simulations predict that when triggered by incident power of 0.13 W, which is just above the triggering threshold, the response time is $~8 \times 10^3$ μs. This response time is reduced by three orders of magnitude, to ~20 μs, by increasing the incident power ten-fold to 1.3 W [Figure S12(b)]. Similarly, for a given incident power [e.g., 0.13 W as shown in Figure S12(c)], a faster response can be expected at a higher bias temperature, which in our experiments we achieve using a bias current.

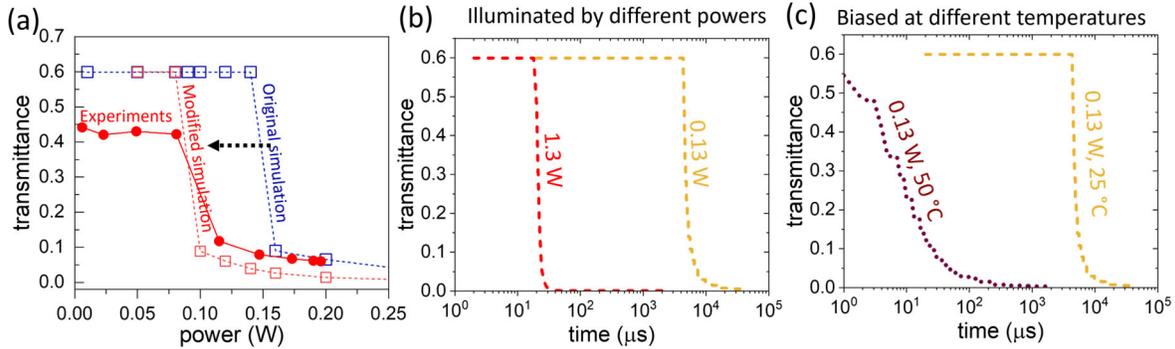

**Figure S12. (a)** Transmittance vs. power during backward illumination for our original optothermal model featured in Figure 4 of the main text (blue dotted curve), modified optothermal model (red dotted curve), and experimental measurements (solid red curve). Our modified model has a slightly reduced $TiO_2$ thermal conductivity and beam waist (by 10% in both cases) to better agree with the experiment. **(b)** Using the modified modeling in "a", we simulated the transient transmittance for incident powers of 0.13 and 1.3 W, for biasing temperature of 25 °C. **(c)** Simulated transient transmittance by backward-incident light with power of 0.13 W, for biasing temperatures of 25 and 50 °C.

Note that the recovery time (the transition from the CLOSED state back to the OPEN state after the light is turned off) may be longer than the triggering response time, owing to both the thermal inertia and the intrinsic thermal hysteresis of $VO_2$. However, that the recovery time may not be particularly significant for limiting and some isolation applications.

## 8. Stabilization of measurements when the $VO_2$ is in an intermediate state

This section provides some additional details of the measurements of Fig. 2c, where transmission spectra were taken using the FTIR and microscope for different levels of current driven through the pads of the device, as well as Fig. 3b, where the device was tested as a limiter.

For Fig. 2c, for each current level, the transmittance was repeatedly measured until two consecutive measurements, taken approximately one minute apart, differed by less than the measurement noise (+/- 0.005). The second of the two consecutive measurements is reported in Fig. 2c. For low and high currents, the measurements reached stable values faster than our ability to record data. However, for currents close to the midpoint of the phase transition, i.e. 180 mA to 220 mA, it took a long time for the transmittance to completely converge to its final value (up to 4 minutes). We expect most of the change in transmittance to occur across a narrow temperature window of a few degrees Celsius or less [for example, note that the simulated transmittance decreases from 0.6 to 0.1 from a 10% change in phase volume fraction; see Fig. S8(b)]. In our measurements, we had no temperature controller: we simply set the current and waited for the temperature to stabilize. The long stabilization time is a result of the combination of the lack of feedback in temperature control combined with the thermal mass of our system and the strong dependence of the transmittance on even sub-degree changes in temperature. We emphasize that these long transmittance stabilization times are not to be mistaken for the practical response time of our device. The long stabilization time is related to the sub-degree convergence of our entire system, including the custom circuit board, to steady state temperature whereas response time is related to the speed at which the $VO_2$ in our device can be driven past its transition point for large input powers and without regard for steady state convergence.

For Fig. 3(b), For each point on each curve, the output power was recorded once readings stabilized to change by less than ~0.1 mW per minute. At intensities leading up to the transition, readings stabilized faster than our ability to measure. Once the transition was initiated, our device would rapidly reach a limiting state due to a thermal feedback effect (as observed elsewhere[S2]) and after the abrupt initial decrease, the transmittance slowly converged asymptotically to its steady state value. Much like in our switching experiment, stabilization time is not to be confused with response time, the latter of which we simulated in S.I. section 7.2.

## 9. Supplementary References

S1. Wan, C. *et al.* On the Optical Properties of Thin-Film Vanadium Dioxide from the Visible to the Far Infrared. *Ann. Phys.* **531**, 1900188 (2019).

S2. Sarangan, A., Ariyawansa, G., Vitebskiy, I. & Anisimov, I. Optical switching performance of thermally oxidized vanadium dioxide with an integrated thin film heater. *Opt. Mater. Express* **11**, 2348 (2021).

S3. Zimmers, A. *et al.* Role of thermal heating on the voltage induced insulator-metal transition in VO2. *Phys. Rev. Lett.* **110**, 1–5 (2013).

S4. Murtagh, O., Walls, B. & Shvets, I. V. Controlling the resistive switching hysteresis in VO2 thin films via application of pulsed voltage. *Appl. Phys. Lett.* **117**, (2020).

S5. Yannopoulos, J. C. *The Extractive Metallurgy of Gold*. (Springer US, 1991). doi:10.1007/978-1-4684-8425-0.

S6. Brahlek, M. *et al.* Opportunities in vanadium-based strongly correlated electron systems. *MRS Commun.* **7**, 27–52 (2017).

S7. Lee, S., Hippalgaonkar, K., Yang, F. & Hong, J. Anomalously low electronic thermal conductivity in metallic vanadium dioxide. *Science* **355**, 371–374 (2017).

S8. Goodfellow. Titanium Dioxide - Titania (TiO2). *AZoM* https://www.azom.com/article.aspx?ArticleID=1179 (2019).

S9. Wan, C. *et al.* Ultrathin Broadband Reflective Optical Limiter. *Laser Photonics Rev.* **15**, 1–8 (2021).

S10.    Wan, C. *et al.* Limiting Optical Diodes Enabled by the Phase Transition of Vanadium Dioxide. *ACS Photonics* **5**, 2688–2692 (2018).